\documentclass[]{elsart}

\usepackage{graphicx}
\usepackage{dcolumn}
\usepackage{amsmath}
\usepackage[]{fontenc}
\usepackage{longtable}

\begin{document}
\begin{frontmatter}
\title{Measurements of $\Xi_c^{+}$ Branching Ratios }
The FOCUS Collaboration\footnote{see http://www-focus.fnal.gov/authors.html for
additional author information.}
\author[ucd]{J.~M.~Link}
\author[ucd]{P.~M.~Yager}
\author[cbpf]{J.~C.~Anjos}
\author[cbpf]{I.~Bediaga}
\author[cbpf]{C.~G\"obel}
\author[cbpf]{J.~Magnin}
\author[cbpf]{A.~Massafferri}
\author[cbpf]{J.~M.~de~Miranda}
\author[cbpf]{I.~M.~Pepe}
\author[cbpf]{E.~Polycarpo}   
% ...
\address[ucd]{University of California, Davis, CA 95616}
\address[cbpf]{Centro Brasileiro de Pesquisas F\'isicas, Rio de Janeiro, RJ, Brasil}
\address[cinv]{CINVESTAV, 07000 M\'exico City, DF, Mexico}
\address[cu]{University of Colorado, Boulder, CO 80309}
\address[fnal]{Fermi National Accelerator Laboratory, Batavia, IL 60510}
\nopagebreak
\address[fras]{Laboratori Nazionali di Frascati dell'INFN, Frascati, Italy I-00044}
\address[ugj]{University of Guanajuato, 37150 Leon, Guanajuato, Mexico} 
\address[ui]{University of Illinois, Urbana-Champaign, IL 61801}
\address[iu]{Indiana University, Bloomington, IN 47405}
\address[korea]{Korea University, Seoul, Korea 136-701}
\address[kp]{Kyungpook National University, Taegu, Korea 702-701}
\address[milan]{INFN and University of Milano, Milano, Italy}
\address[nc]{University of North Carolina, Asheville, NC 28804}
\address[pavia]{Dipartimento di Fisica Nucleare e Teorica and INFN, Pavia, Italy}
\address[pr]{University of Puerto Rico, Mayaguez, PR 00681}
\address[sc]{University of South Carolina, Columbia, SC 29208}
\address[ut]{University of Tennessee, Knoxville, TN 37996}
\address[vu]{Vanderbilt University, Nashville, TN 37235}
\address[wisc]{University of Wisconsin, Madison, WI 53706}

\date{\today}
\begin{abstract}
{\normalsize Using data collected by the fixed target Fermilab experiment FOCUS, we
measure the branching ratios of the Cabibbo favored decays $\Xi_c^+ 
\rightarrow \Sigma^+K^-\pi^+$, $\Xi_c^+ \rightarrow \Sigma^+ \bar{K}^{*}(892)^0$, 
and $\Xi_c^+ \rightarrow \Lambda^0K^-\pi^+\pi^+$ relative to $\Xi_c^+ \rightarrow \Xi^-\pi^+\pi^+$ 
to be $0.91\pm0.11\pm0.04$, $0.78\pm0.16\pm0.06$, and $0.28\pm0.06\pm0.06$, 
respectively. We report the first observation 
of the Cabibbo suppressed decay $\Xi_c^+ \rightarrow \Sigma^+K^+K^-$ and
we measure the branching ratio relative to 
$\Xi_c^+ \rightarrow \Sigma^+K^-\pi^+$ to be $0.16\pm0.06\pm0.01$. We also set
90\% confidence level upper limits
for $\Xi_c^+ \rightarrow \Sigma^+ \phi$ and
$\Xi_c^+ \rightarrow \Xi^*(1690)^0(\Sigma^+ K^-) K^+$ relative to 
$\Xi_c^+ \rightarrow \Sigma^+K^-\pi^+$ to be 0.12 and 0.05, respectively.
We find an indication of the decays
$\Xi_c^+ \rightarrow \Omega^-K^{+}\pi^+$ 
and $\Xi_c^+ \rightarrow \Sigma^{*}(1385)^+ \bar{K}^0$ 
and set 90\% confidence level upper limits for the branching ratios
with respect to $\Xi_c^+ \rightarrow \Xi^-\pi^+\pi^+$
to be 0.12 and 1.72, respectively.
Finally, we determine the 90\% C.L. upper limit for the 
resonant contribution $\Xi_c^+ \rightarrow \Xi^{*}(1530)^0 \pi^+$ relative 
to $\Xi_c^+ \rightarrow \Xi^-\pi^+\pi^+$ to be 0.10.
} 
{\normalsize \par}
\end{abstract}

\end{frontmatter}

\section{Introduction}

\begin{figure}[!]
\begin{center}\includegraphics[  height=12cm ]{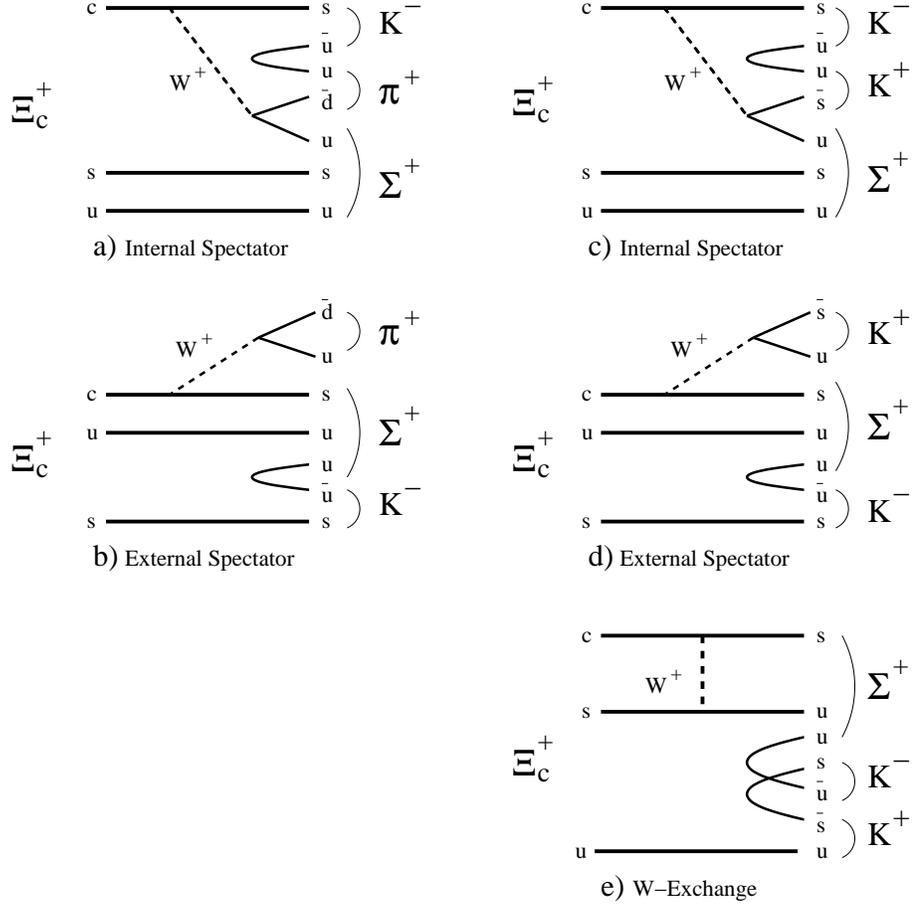}\end{center}
\caption{\label{diagrams} Possible weak diagrams for a) and b): Cabibbo favored decay 
$\Xi_c^+\rightarrow \Sigma^+ K^{-}\pi ^{+}$; c), d), e): Cabibbo suppressed decay 
$\Xi_c^+\rightarrow \Sigma^+ K^{+}K ^{-}$. The W-exchange diagram contributes only
to the Cabibbo suppressed decay.}
\end{figure}

In addition to several improved measurements of $\Xi_c^+$ branching ratios, 
we report an indication of new $\Xi_c^+$ decay modes and the first observation of the
Cabibbo suppressed decay $\Xi _{c}^{+}\rightarrow \Sigma^+K^+K^-$.
These analyses may provide useful information about the various charm baryon weak decay
mechanisms. 
In particular we find a suggestion of the decay 
$\Xi _{c}^{+}\rightarrow \Sigma^*(1385)^+ \bar{K}^0$
for which flavor symmetry arguments predict
a zero amplitude~\cite{Korner}. A non-vanishing amplitude could be related 
to spin-spin interactions between the light quarks in the baryon $\Xi_c^+$ \cite{Hussain}.
As regards the $\Xi _{c}^{+}\rightarrow \Sigma^+K^+K^-$, we measure the branching ratio relative to
the Cabibbo favored mode $\Xi _{c}^{+}\rightarrow \Sigma^+K^-\pi^+$.
While tree diagrams (internal and external spectator) contribute to both 
Cabibbo favored and Cabibbo suppressed modes, the W-exchange diagram 
contributes only to the Cabibbo suppressed decay (Fig.~\ref{diagrams}). 
Assuming a similar contribution from strong interactions for the two modes,
and neglecting possible resonant structure,
one might naively extract information on the role of the W-exchange diagram. 
This result may also aid in understanding the discrepancy between the predicted and measured 
$\Xi _{c}^{+}$ lifetime~\cite{Guberina:2002fz,Link:2001qy}.
 
\section{Event Reconstruction}

FOCUS is a photoproduction experiment which collected data during the
1996--1997 fixed-target run at Fermilab. The apparatus is equipped
with precise vertex and comprehensive particle identification detectors. 
For about 2/3 of the data taking a $25~\mu\textrm{m}$ pitch silicon strip detector
(TS)~\cite{TS_desc} was interleaved with the BeO target segments.
The spectrometer is divided into an inner region for high momentum
track reconstruction and an outer region for low momentum tracks. 

All decay modes reported have a hyperon in the final state. The $\Sigma ^{+}$
particles are reconstructed in both $p$$\pi ^{0}$ and
$n$$\pi ^{+}$ decay modes. As the direction of the neutral particle is not reconstructed,
kinematic constraints are used to compute the $\Sigma ^{+}$ momentum.
If the decay occurs upstream of the magnetic field, there is
a two-fold ambiguity in the $\Sigma ^{+}$ momentum. The $\Xi ^{-}$
and $\Omega ^{-}$ are reconstructed in the modes $\Lambda ^{0}$$\pi ^{-}$
and $\Lambda ^{0}$$K^{-}$, respectively, while $\Lambda ^{0}$ decays
are reconstructed in the charged mode\footnote{Throughout this paper the charged 
conjugate decay is understood.} $p$$\pi ^{-}$. A detailed description of the hyperon reconstruction techniques
in FOCUS is reported in Reference~\cite{Link:2001dj}.

Candidates are reconstructed by first forming a vertex with tracks
consistent with a specific charm decay hypothesis. A cut on the confidence
level (CLD) that these tracks form a good vertex is applied. The production
vertex is found using a candidate driven vertex algorithm
which uses the final state momentum to define the line of flight of
the charm particle~\cite{Driv_alg}. The seed track for the charm
particle is used to form a production vertex with at least two other
tracks in the target region. We require a value of at least $1\%$
for the confidence level of the production vertex. Most of the
background is rejected by applying a separation cut between the production
and decay vertices (we require the significance of
separation, $L/\sigma _{L}$, between the two vertices to be greater
than some number). \v{C}erenkov identification~\cite{Link:2001pg} is required
on each charged final state particle in the decay. For each hypothesis
($\alpha =$ electron, pion, kaon or proton) we construct a $\chi ^{2}$-like
variable $W_{\alpha }=-2 \log$~(likelihood). We use either a requirement
that one hypothesis, $\beta$, is favored with respect to another hypothesis,
$\alpha$, ($W_{\alpha }-W_{\beta }>n)$ or a requirement that one
hypothesis is favored with respect to all the other
hypotheses ($\min\{W_{\alpha }\}-W_{\beta }>n$).

In order to minimize systematic biases, the normalization mode is selected
using the same cuts as the specific decay when possible. Differences  between
each mode and its reference mode will be discussed below. 
The evaluation of efficiencies accounts for the decay
fractions of the observed daughters.

\section{$\Xi _{c}^{+}$ decays containing a $\Sigma ^{+}$ particle}

We measure the branching ratio of
$\Xi _{c}^{+}\rightarrow \Sigma ^{+}K^{-}\pi ^{+}$
and $\Xi _{c}^{+}\rightarrow \Sigma ^{+}\bar{K}^{*}(892)^{0}$ relative
to $\Xi _{c}^{+}\rightarrow \Xi ^{-}\pi ^{+}\pi ^{+}$. The decay
mode $\Xi _{c}^{+}\rightarrow \Sigma ^{+}K^{-}\pi ^{+}$ 
is selected by requiring $\mathrm{CLD}>1\%$
while for $\Xi _{c}^{+}\rightarrow \Xi ^{-}\pi ^{+}\pi ^{+}$ we
require $\mathrm{CLD}>2\%$. A 
minimum cut of $40~\textrm{GeV}/c$ is applied on the $\Xi _{c}^{+}$ momentum.
Due to different levels of background, we require $L/\sigma _{L}>9.5$
for $\Xi _{c}^{+}\rightarrow \Sigma ^{+}K^{-}\pi ^{+}$ and $L/\sigma _{L}>4.5$
for $\Xi _{c}^{+}\rightarrow \Xi ^{-}\pi ^{+}\pi ^{+}$. Each pion
from the charm decay vertex must satisfy $\min\{W_{\alpha }\}-W_{\pi }>-6$.
In the $\Xi _{c}^{+}\rightarrow \Sigma ^{+}K^{-}\pi ^{+}$ mode the
kaon hypothesis must be favored over the pion hypothesis ($W_{\pi }-W_{K}>1$).
To eliminate possible contamination from the $\Lambda _{c}^{+}\rightarrow \Sigma ^{+}\pi ^{+}\pi ^{-}$
decays, where the $\pi ^{-}$ is misidentified as a $K^{-}$, we increase
the $K-\pi $ separation cut from 1 to 5 for those events which, reconstructed
as $\Sigma ^{+}\pi ^{+}\pi ^{-}$, fall within $30~\textrm{MeV}/c^2$
of the nominal $\Lambda _{c}^{+}$ mass. A loose requirement, $W_{p}-W_{\pi }>-3$,
is applied on proton-pion separation. In addition, we
reject candidates with a decay proper time resolution ($\sigma_t$) less than $110$~fs ($140$~fs) for TS (not
TS) run period events. Further, a muon incompatibility cut is imposed
on the kaon and pion for $\Xi _{c}^{+}\rightarrow \Sigma ^{+}K^{-}\pi ^{+}$ candidates. 

\begin{figure}[!]
\begin{center}\includegraphics[  width=6cm,
  height=8cm]{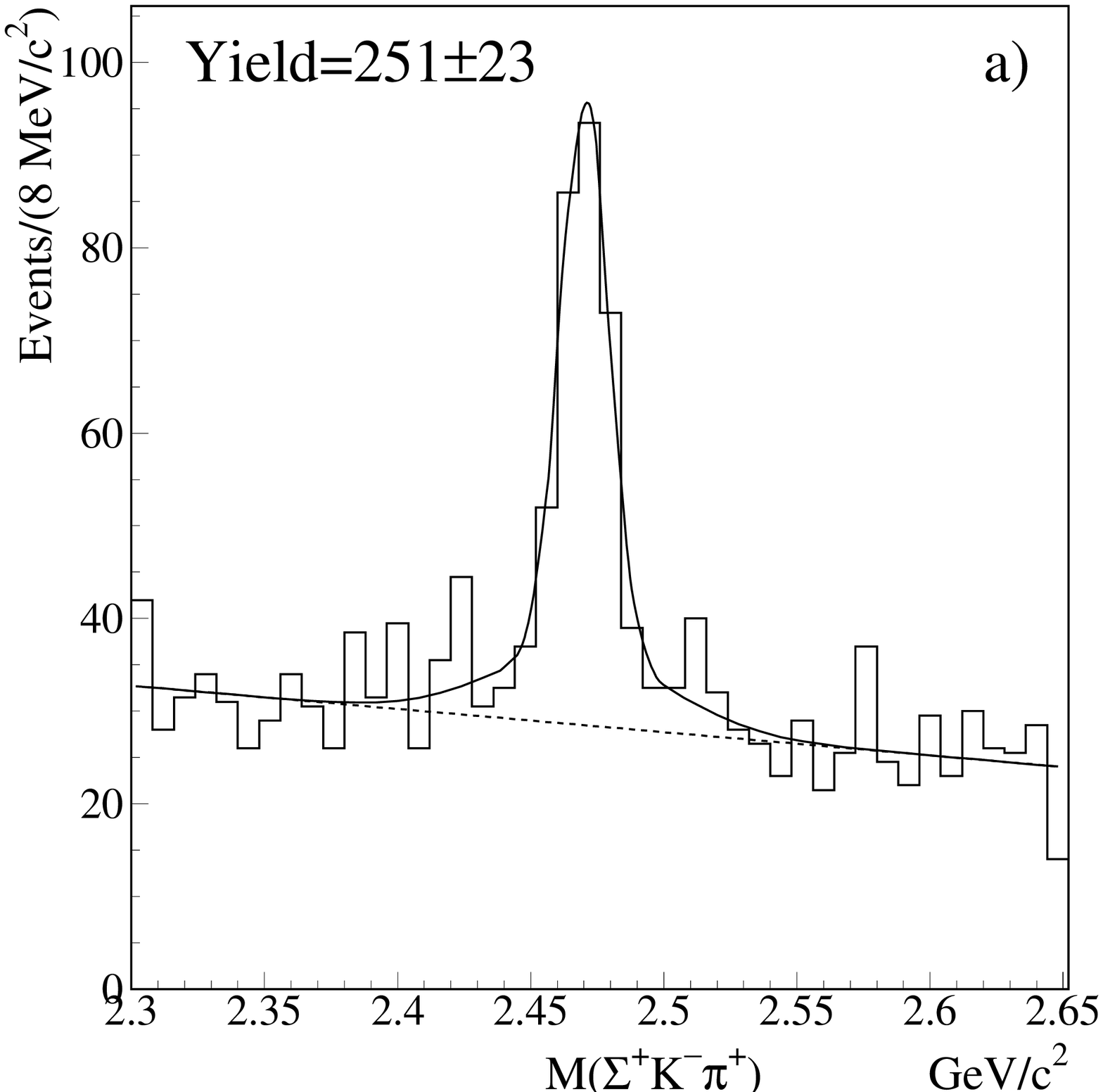}\includegraphics[  width=6cm,
  height=8cm]{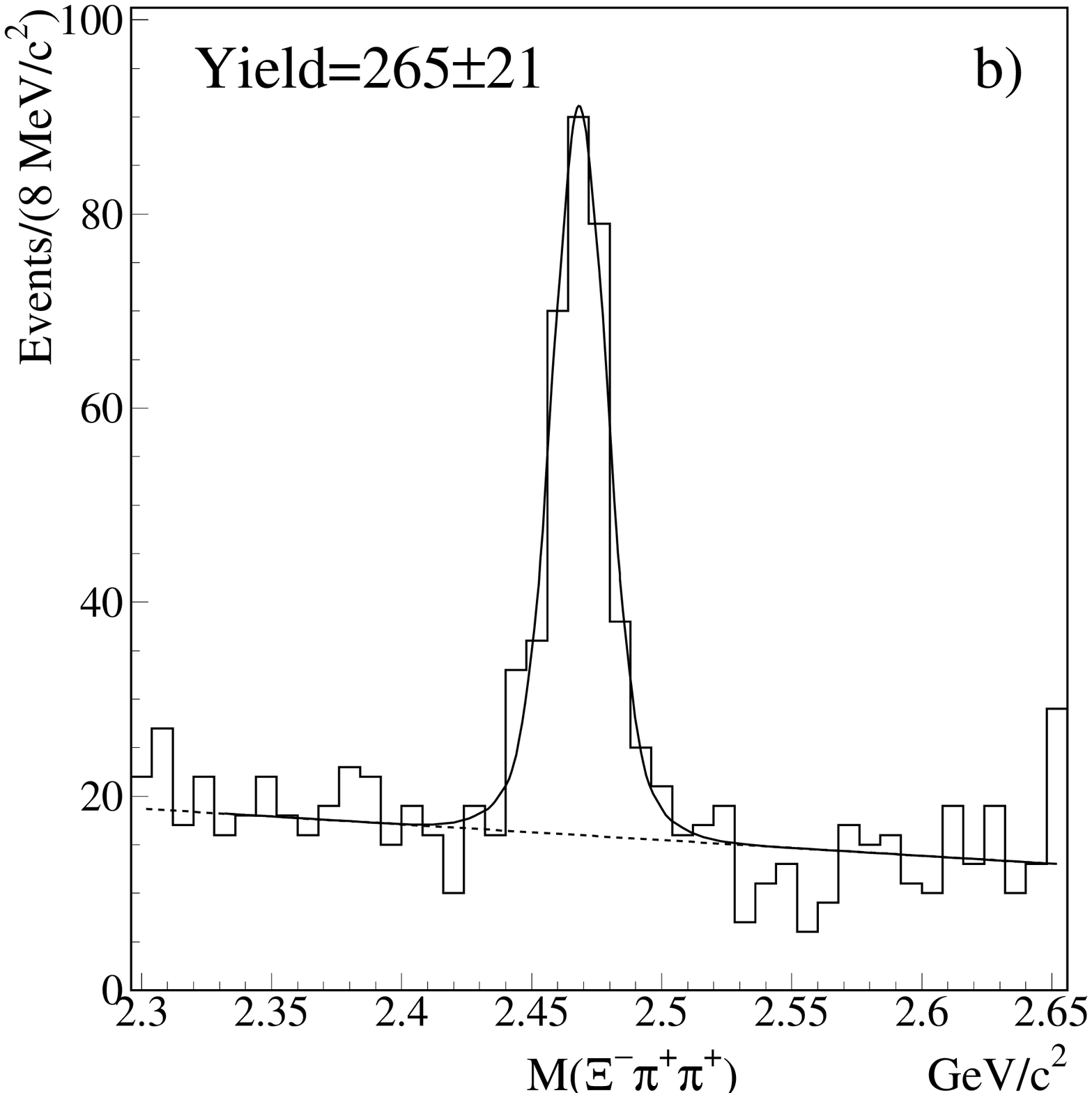}\end{center}

\caption{\label{cap:skpi and cascpipi}Invariant mass distribution of: a)
$\Xi _{c}^{+}\rightarrow \Sigma ^{+}K^{-}\pi ^{+}$. 
b) $\Xi _{c}^{+}\rightarrow \Xi ^{-}\pi ^{+}\pi ^{+}$. For both modes 
the fit has been performed using two Gaussians for the signal and a first order polynomial 
for the background.}
\end{figure}
In Fig. \ref{cap:skpi and cascpipi} the invariant mass distributions
for $\Sigma ^{+}K^{-}\pi ^{+}$ and $\Xi ^{-}\pi ^{+}\pi ^{+}$ are
presented. A good fit function to our data is two Gaussian distributions for the signal 
and a first order polynomial for the background, especially for decays with a
two-fold ambiguity. For the $\Sigma ^{+}K^{-}\pi ^{+}$ mode
the fit returns a yield of $251\pm 23$ events.
For this mode, the sigmas and the ratio of the yields of the two Gaussians, and the
mean of the wide Gaussian are fixed to the Monte Carlo values. The $\Xi ^{-}\pi ^{+}\pi ^{+}$
distribution is also fit using two Gaussians for the signal and
a first order polynomial for the background. The resultant yield
is $265\pm 21$ events. A Monte Carlo simulation is used to determine
the relative efficiency. We find no significant change in the
$\Xi _{c}^{+}\rightarrow \Sigma ^{+}K^{-}\pi ^{+}$ efficiency due to the 
$\Xi _{c}^{+}\rightarrow \Sigma ^{+}K^{*}(892)^0$ contribution.
We determine the branching ratio to be
\begin{equation}
\frac{\Gamma (\Xi _{c}^{+}\rightarrow \Sigma ^{+}K^{-}\pi ^{+})}{\Gamma (\Xi _{c}^{+}
\rightarrow \Xi ^{-}\pi ^{+}\pi ^{+})}=0.91\pm 0.11~\textrm{(stat)}.
\end{equation}
For the $\Xi _{c}^{+}\rightarrow \Sigma ^{+}\bar{K}^{*}(892)^{0}$
mode we fit the $K^{-}\pi ^{+}$ invariant mass distribution.
We select events in the $\Sigma ^{+}K^{-}\pi ^{+}$ signal region
(mass window within $30~\textrm{MeV}/c^2$ of the fit mass),
and subtract events in the sidebands (two symmetric regions
$70~\textrm{MeV}/c^2$ to $100~\textrm{MeV}/c^2$ away from the fit mass).
The $\Xi _{c}^{+}\rightarrow \Sigma ^{+}\bar{K}^{*}(892)^{0}$
events are selected with the same selection cuts as those used in the $\Xi _{c}^{+}\rightarrow \Sigma ^{+}K^{-}\pi ^{+}$
branching ratio measurement. The $K^{-}\pi ^{+}$ invariant mass distribution
is fit using a Breit-Wigner (with width fixed to the Monte Carlo value)
for the signal and the non-resonant
$\Xi _{c}^{+}\rightarrow \Sigma ^{+}K^{-}\pi ^{+}$ shape determined
with the Monte Carlo simulation. 
\begin{figure}[!]
\begin{center}\includegraphics[  width=8cm,
  height=8cm]{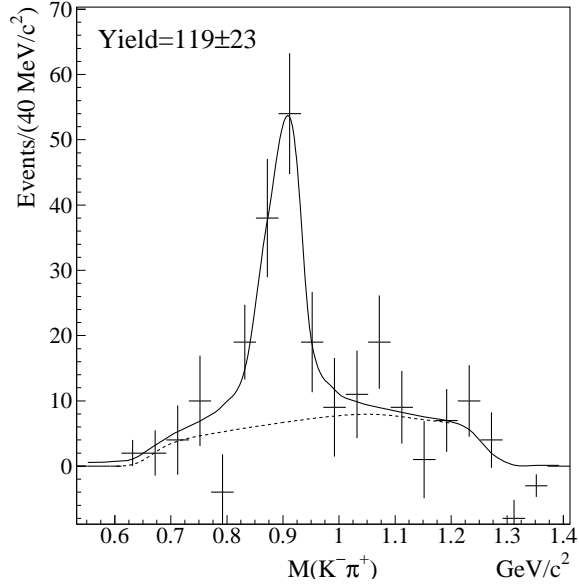}\end{center}

\caption{\label{kstar}$K^{-}\pi ^{+}$ invariant mass distribution (sideband
subtracted). The fit is performed using a Breit-Wigner distribution
for the signal and a shape for the $\Xi _{c}^{+}\rightarrow \Sigma ^{+}K^{-}\pi ^{+}$
non-resonant component taken from a high statistics Monte Carlo simulation. The width of the
Breit-Wigner is fixed to the Monte Carlo value.}
\end{figure}
In Fig. \ref{kstar} we present the $K^{-}\pi ^{+}$ invariant mass
distribution after sideband subtraction. The yield is $119\pm 23$
events. The resulting branching ratio relative to $\Xi _{c}^{+}\rightarrow \Xi ^{-}\pi ^{+}\pi ^{+}$ is
\begin{equation}
\frac{\Gamma (\Xi _{c}^{+}\rightarrow \Sigma ^{+}\bar{K}^{*}(892)^{0})}{\Gamma
(\Xi _{c}^{+}\rightarrow \Xi ^{-}\pi ^{+}\pi ^{+})}= 0.78\pm 0.16~\textrm{(stat)}.
\end{equation}

We report the first observation of the Cabibbo suppressed decay $\Xi _{c}^{+}\rightarrow \Sigma ^{+}K^{-}K^{+}$
and measure the branching ratio with respect to the similar mode $\Xi _{c}^{+}\rightarrow \Sigma ^{+}K^{-}\pi ^{+}$.
Due to the larger level of background and lower efficiency for the
$\Xi _{c}^{+}\rightarrow \Sigma ^{+}(n\pi ^{+})K^{-}K^{+}$ mode,
we only use the signal from $\Xi _{c}^{+}\rightarrow \Sigma ^{+}(p\pi ^{0})K^{-}K^{+}$
decays. To minimize possible systematic biases, we restrict the normalizing
mode to events in which the $\Sigma ^{+}$ decays via $p$$\pi ^{0}$.
The selection cuts used to select this sample are similar to
the cuts used in the inclusive $\Sigma ^{+}K^{-}\pi ^{+}$ mode. The
main differences are the $\Xi _{c}^{+}$ minimum momentum cut, which is reduced
to $30~\textrm{GeV}/c$, and the $L/\sigma _{L}$ cut, which is reduced to
8.5. 
To eliminate contamination from $\Xi_c^+ \rightarrow \Sigma^+ K^-
\pi^+$ events, $\Sigma^+ K^+ K^-$ candidates which, when reconstructed
as $\Sigma^+ K^- \pi^+$, fall near the $\Xi_c^+$ mass, are eliminated.
\begin{figure}[!]
\begin{center}\includegraphics[  width=8cm,
  height=8cm]{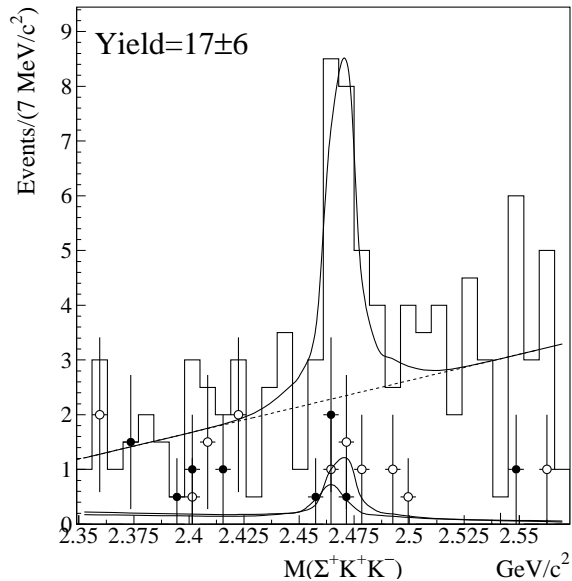}\end{center}

\caption{\label{skk}The histogram shows the inclusive $\Sigma ^{+}(p\pi ^{0})K^{+}K^{-}$ invariant
mass distribution, the data is fit to two Gaussians for the signal and a first order polynomial
for the background. The points with error bars show the possible contribution from
$\Sigma ^{+}\phi$ (empty circles) and $\Xi^*(1690)^0 K^{+}$ (filled circles).}
\end{figure}
 The $\Sigma ^{+}K^{+}K^{-}$ invariant mass distribution is shown
in Fig. \ref{skk}. The fit is performed using a double Gaussian for
the signal and a first order polynomial for the background. Again,
the ratio of yields, the resolutions of the two Gaussians and the mean
of the wide Gaussian are fixed to the Monte Carlo values. The fit
returns $17\pm 6$ events. The branching ratio relative 
to $\Xi _{c}^{+}\rightarrow \Sigma ^{+}K^{-}\pi ^{+}$ is
\begin{equation}
\frac{\Gamma (\Xi _{c}^{+}\rightarrow \Sigma ^{+}K^{+}K^{-})}{\Gamma (\Xi _{c}^{+}\rightarrow \Sigma
^{+}K^{-}\pi ^{+})}=0.16\pm 0.06~\textrm{(stat).}
\end{equation}
As significant resonant structure is observed in the decay 
$\Lambda_c^+ \rightarrow \Sigma^+ K^+ K^-$~\cite{Lore,Belle}, we search for possible 
contribution from $\Xi_c^+ \rightarrow \Sigma^+ \phi$ and $\Xi_c^+ \rightarrow \Xi^*(1690)^0 K^+$.
For both decays we fit the $\Sigma^+ K^+ K^-$ invariant mass distribution.
For $\Xi_c^+ \rightarrow \Sigma^+ \phi$ decay we make a sideband subtraction on the $K^+ K^-$ invariant
mass (using $20~\textrm{MeV}/c^2$ wide signal region and sideband). 
For $\Xi_c^+ \rightarrow \Xi^*(1690)^0 K^+$ we require the $\Sigma^+ K^-$ invariant mass to
be within $20~\textrm{MeV}/c^2$ of the nominal $\Xi^*$ mass (where
we assume no contribution from the non-resonant mode), and we exclude events in the
$\phi$ signal region. No significant contribution is found. In Fig.~\ref{skk} we show the fits of 
the two resonant modes superimposed to the inclusive
sample. The fit reports $3\pm2$ events for $\Sigma^+ \phi$ and $2\pm2$ for $\Xi^*(1690)^0 K^+$.
We set the upper limit at 90\% confidence level
for the branching fractions relative to $\Xi_c^+\rightarrow \Sigma^+ K^- \pi^+$ to be
\begin{equation}
\frac{\Gamma (\Xi_c^+ \rightarrow \Sigma^+ \phi)}
{\Gamma (\Xi _{c}^{+}\rightarrow \Sigma ^{+}K ^{-}\pi^+)}<0.12
\end{equation} 
and
\begin{equation}
\frac{\Gamma (\Xi_c^+ \rightarrow \Xi(1690)^0 K^+)}
{\Gamma (\Xi _{c}^{+}\rightarrow \Sigma ^{+}K ^{-}\pi^+)}<0.05,
\end{equation} 
where no correction is made for the branching ratio of $\Xi^*(1690)^0 \rightarrow \Sigma^+ K^-$.
For both modes we find a negligible systematic uncertainty. 
\section{$\Xi _{c}^{+}\rightarrow \Lambda ^{0}K^{-}\pi ^{+}\pi ^{+}$, $\Xi _{c}^{+}\rightarrow \Omega ^{-}K^{+}\pi ^{+}$
and $\Xi _{c}^{+}\rightarrow \Sigma ^{*}(1385)^+ \bar{K}^{0}$ decays}
\begin{figure}[!]
\begin{center}\includegraphics[  width=8cm,
  height=8cm]{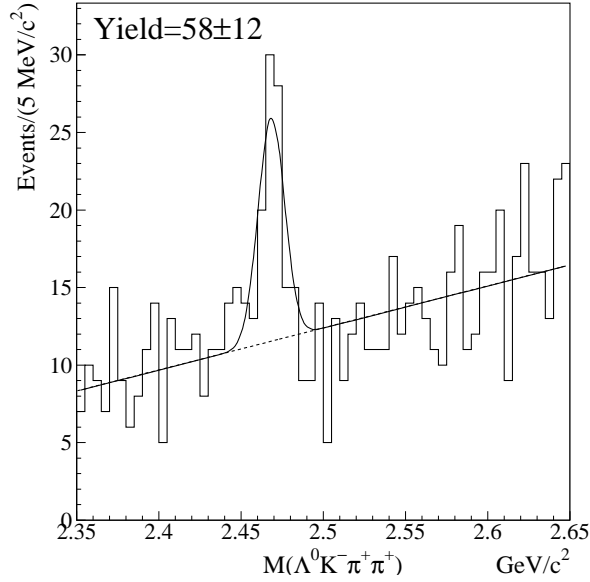}\end{center}
\caption{\label{lkpp}Invariant mass distribution for $\Lambda ^{0}K^{-}\pi ^{+}\pi ^{+}$.
The fit function is a sum of a Gaussian for the signal and a linear
background.}
\end{figure}
We measure the branching ratio of the decay $\Xi _{c}^{+}\rightarrow \Lambda ^{0}K^{-}\pi ^{+}\pi ^{+}$
relative to $\Xi _{c}^{+}\rightarrow \Xi ^{-}\pi ^{+}\pi ^{+}$. The
sample is selected requiring a significance of separation ($L/\sigma _{L}$)
greater than $5$, $\mathrm{CLD}>2\%$, and $\sigma_t<100$~fs. Furthermore, the kaon
hypothesis must be favored over the pion hypothesis ($W(\pi )-W(K)>2$),
while the pion must satisfy $\min\{W_{\alpha }\}-W_{\pi }>-6$.
The invariant mass distribution for $\Lambda ^{0}K^{-}\pi ^{+}\pi ^{+}$
is shown in Fig.~\ref{lkpp}. The fit is performed using a Gaussian
for the signal plus a linear polynomial for the background. The signal
yield is $58\pm 12$ events. The same selection cuts are applied to
the normalization mode $\Xi _{c}^{+}\rightarrow \Xi ^{-}\pi ^{+}\pi ^{+}$
to minimize possible systematic biases. We find the branching ratio of
$\Xi _{c}^{+}\rightarrow \Lambda ^{0}K^{-}\pi ^{+}\pi ^{+}$
relative to $\Xi _{c}^{+}\rightarrow \Xi ^{-}\pi ^{+}\pi ^{+}$ to
be
\begin{equation}
\frac{\Gamma (\Xi _{c}^{+}\rightarrow \Lambda ^{0}K^{-}\pi ^{+}\pi ^{+})}
{\Gamma (\Xi _{c}^{+}\rightarrow \Xi ^{-}\pi ^{+}\pi ^{+})}=0.28\pm 0.06~\textrm{(stat)}.
\end{equation}

We find an indication of the decay $\Xi _{c}^{+}\rightarrow \Omega ^{-}K^{+}\pi ^{+}$.
The sample is selected by reconstructing the $\Omega ^{-}$ when it
decays to $\Lambda ^{0}K^{-}$. The $\Lambda ^{0}K^{-}$ invariant
mass must be within $20~\textrm{MeV}/c^2$ of the nominal $\Omega ^{-}$ 
mass and the decay vertex must satisfy a minimum confidence
level cut of $1\%$. The significance of separation, $L/\sigma _{L}$,
must be greater than 0.5. The kaon from the decay vertex must be favored
with respect to the pion hypothesis ($W(\pi )-W(K)>2$), while
the pion must satisfy $\min\{W_{\alpha }\}-W_{\pi }>-6$\@. 
\begin{figure}[!]
\begin{center}\includegraphics[  width=8cm,
  height=8cm]{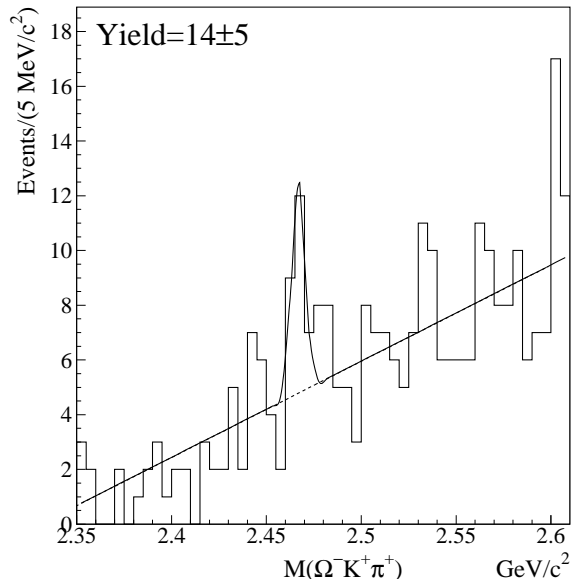}\end{center}
\caption{\label{okpi}Invariant mass distribution for the combination $\Omega ^{-}K^{+}\pi ^{+}$.
The fit is performed using a single
Gaussian for the signal plus a first order polynomial for the
background.}
\end{figure}
The $\Omega ^{-}K^{+}\pi ^{+}$ invariant mass distribution is shown in Fig.~\ref{okpi}\@. The data is fit
with a single Gaussian for the signal and a linear polynomial for
the background. We used similar cuts for the normalization mode. We
report the value, for the branching ratio of $\Xi _{c}^{+}\rightarrow \Omega ^{-}K^{+}\pi ^{+}$
relative to $\Xi _{c}^{+}\rightarrow \Xi ^{-}\pi ^{+}\pi ^{+}$, to
be
\begin{equation}
\frac{\Gamma (\Xi _{c}^{+}\rightarrow \Omega ^{-}K^{+}\pi ^{+})}
{\Gamma (\Xi _{c}^{+}\rightarrow \Xi ^{-}\pi ^{+}\pi ^{+})}=0.07\pm 0.03~\textrm{(stat)}.
\end{equation}
After evaluation of the systematic uncertainty as described in the last section, we measure the upper limit
at 90\% confidence level to be
\begin{equation}
\frac{\Gamma (\Xi _{c}^{+}\rightarrow \Omega ^{-}K^{+}\pi ^{+})}
{\Gamma (\Xi _{c}^{+}\rightarrow \Xi ^{-}\pi ^{+}\pi ^{+})}<0.12.
\end{equation}

We also see an indication of the decay
$\Xi_c^+ \rightarrow \Sigma^*(1385)^+\bar{K}^0$ where the $\Sigma^*$ is reconstructed
in the decay mode $\Lambda ^{0}$$\pi ^{+}$.
The invariant mass of this combination is required to be in the interval
$1.349$--$1.421~\textrm{GeV}/c^2$ which corresponds to a $\pm 1.0~\Gamma$ window around the
$\Sigma ^{*}$ nominal mass. The $\bar{K}^{0}$ is reconstructed as
a $K_{S}^{0}$ in the $\pi ^{+}$$\pi ^{-}$ decay mode. We require that the reconstructed invariant mass of
the $\pi^+\pi^-$ lie within 3 standard deviations of the nominal $K^0_S$ mass.
We select the events by requiring $\mathrm{CLD}>3\%$ and the significance
of detachment $L/\sigma _{L}$ greater than 4.5. We
also reject events where the $\pi^+$ track from the decay vertex has a confidence
level greater than 0.1\% of coming from the production vertex. Further,
the $\Xi _{c}^{+}$ candidates must have a momentum greater than $45~\textrm{GeV}/c$.
We identify the pion from the $\Sigma ^{*}$ by requiring $\min \{W_{\alpha }\}-W_{\pi }>-6$\@.
\begin{figure}[!]
\begin{center}\includegraphics[  width=8cm,
  height=8cm]{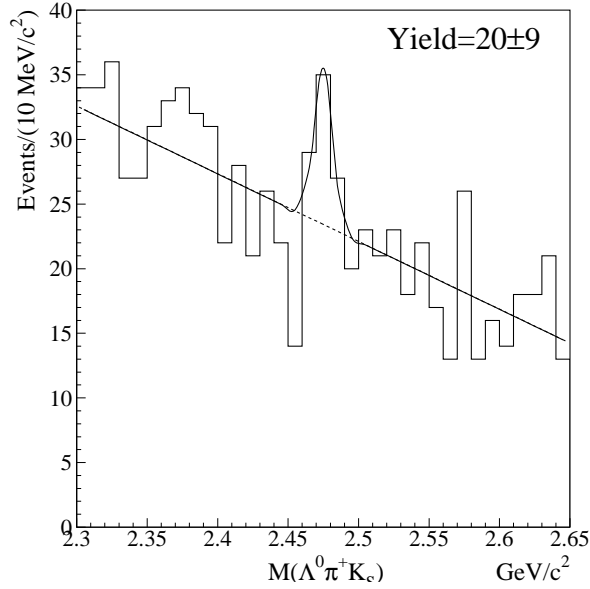}\end{center}
\caption{\label{sigmastark}Invariant mass of the $\Lambda ^{0}\pi ^{+}K_{S}^{0}$ combination 
for the $\Xi _{c}^{+}\rightarrow \Sigma ^{*}(1385)^+ \bar{K}^{0}$ decay mode.
The fit is to a Gaussian for the signal events and a first order polynomial
for the background. }
\end{figure}
In Fig.\ref{sigmastark} the $\Lambda ^{0}\pi ^{+}K_{S}^{0}$ invariant
mass is shown. We measure the branching ratio relative to $\Xi _{c}^{+}\rightarrow \Xi ^{-}\pi ^{+}\pi ^{+}$
to be
\begin{equation}
\frac{\Gamma (\Xi _{c}^{+}\rightarrow \Sigma ^{*}(1385)^+ \bar{K}^{0})}{\Gamma (\Xi _{c}^{+}\rightarrow
\Xi ^{-}\pi ^{+}\pi ^{+})}=1.00\pm 0.49~\textrm{(stat)}.
\end{equation}
We find the upper limit for the branching ratio at 90\% confidence level to be
\begin{equation}
\frac{\Gamma (\Xi _{c}^{+}\rightarrow \Sigma ^{*}(1385)^+ \bar{K}^{0})}{\Gamma (\Xi _{c}^{+}\rightarrow
\Xi ^{-}\pi ^{+}\pi ^{+})}<1.72;
\end{equation}
this measurement includes the systematic uncertainty.

\section{Search for the resonant decay $\Xi_c^+ \rightarrow \Xi^*(1530)^0\pi^+$}
\begin{figure}[!]
\begin{center}\includegraphics[  width=8cm,
  height=8cm]{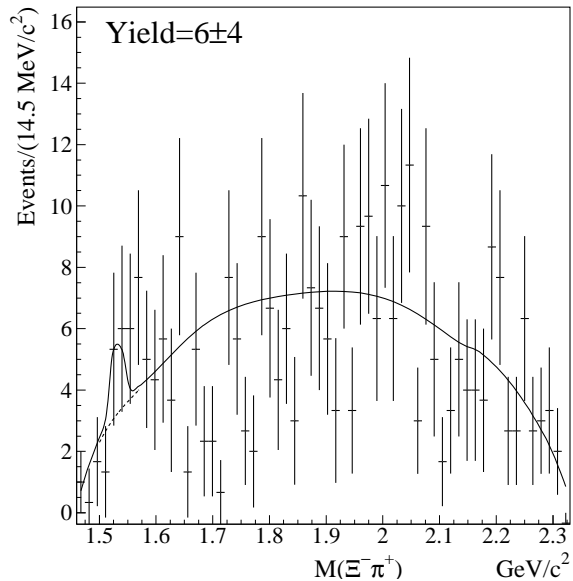}\end{center}
\caption{\label{cascstar} A fit to the $\Xi ^{-}\pi ^{+}$ sideband-subtracted
invariant mass distribution performed using a Breit-Wigner for the
signal region plus a shape for the non-resonant $\Xi _{c}^{+}\rightarrow \Xi ^{-}\pi ^{+}\pi ^{+}$
and the wrong ($\Xi ^{-}$$\pi ^{+}$) combinations taken by Monte
Carlo simulation. The Breit-Wigner width and mean are fixed to the Monte Carlo values.}
\end{figure}

As most of the branching ratios are computed relative to $\Xi _{c}^{+}\rightarrow \Xi ^{-}\pi ^{+}\pi^{+}$,
we investigate possible systematic errors due to a contribution from $\Xi _{c}^{+}\rightarrow \Xi^{*}(1530)^0 \pi ^{+}$.
The decay width of this mode is expected to be zero~\cite{Korner}. 
In Fig. \ref{cascstar} we plot the sideband subtracted invariant mass distribution for the
two possible combinations of $\Xi ^{-}\pi ^{+}$ in the $\Xi^- \pi^+ \pi^+$ sample. We fit the signal
events using a Breit-Wigner. The background is given by two contributions, the non-resonant
$\Xi _{c}^{+}\rightarrow \Xi ^{-}\pi ^{+}\pi ^{+}$ events and the
wrong $\Xi ^{-}\pi ^{+}$ combination. Both shapes for these
distributions are obtained from a Monte Carlo simulation. 
The width and mean of the Breit-Wigner and the ratio between the Breit-Wigner amplitude and
the amplitude of the wrong sign combination, are fixed to the Monte Carlo
values.
No significant contribution from this resonant structure is found. After evaluation of 
the systematic uncertainty, we find the upper limit at $90\%$ confidence level for the 
branching ratio
relative to $\Xi_c^+ \rightarrow \Xi ^{-}\pi ^{+}\pi ^{+}$ to be
\begin{equation}
\frac{\Gamma (\Xi _{c}^{+}\rightarrow \Xi ^{*}(1530)^0 \pi ^{+})}
{\Gamma (\Xi _{c}^{+}\rightarrow \Xi ^{-}\pi ^{+}\pi ^{+})}<0.10.
\end{equation} 
We calculate that in the case of a contamination from the resonant substructure up to a level of 10\%, 
the efficiency of $\Xi^-\pi^+\pi^+$ inclusive would change by less than 1\%.
For this reason the $\Xi^- \pi^+ \pi^+$ efficiencies for the branching ratio measurements
have been evaluated with a non-resonant Monte Carlo.

\section{Systematic studies}

The systematic uncertainties are evaluated after investigation of two possible sources:
the choice of fitting conditions and the Monte Carlo simulation.
The total systematic error is computed by adding in quadrature these two independent contributions. 
We measure the systematic uncertainty due to fitting conditions using
a fit variation technique, which includes variations in bin size, fitting range, 
background shapes, sidebands size and position. 
To assess possible systematic uncertainties related to the Monte Carlo simulation we used 
the standard FOCUS split sample technique, described in~\cite{annalisa}, and based on 
the S-factor method used by the Particle Data Group~\cite{pdg}. We investigate possible biases due to
poor simulation of variables such as run period, particle and
antiparticle, $\Sigma ^{+}$ decay mode and momentum, $\Xi _{c}^{+}$
momentum and significance of separation between production and decay
vertices. 
Furthermore, as noted above, we find that the efficiency of the $\Xi^-\pi^+\pi^+$
mode is not affected by possible resonant structure.
Due to the low statistics, no split sample studies are made for $\Xi _{c}^{+}\rightarrow \Sigma
^{+}K^{+}K^{-}$, 
$\Xi _{c}^{+}\rightarrow \Sigma ^{+}\phi$,
$\Xi _{c}^{+}\rightarrow \Xi^*(1690)^0 K^{+}$,
$\Xi _{c}^{+}\rightarrow \Xi ^{*}(1530)^- \pi^+$
and $\Xi _{c}^{+}\rightarrow \Sigma ^{*}(1385)^+ \bar{K}^{0}$. Because of
the particular spin properties of the particles involved in the
latter decay mode, we evaluated a possible systematic uncertainty 
of our simulation by varying the Monte Carlo angular distribution to match the shape
obtained in the data. 
In Table \ref{system} we summarize the systematic uncertainty for each mode.
In Table \ref{tb:results} we present the FOCUS results with a comparison
to previous measurements from CLEO~\cite{Bergfeld:1996qn} and SELEX~\cite{Jun:00}.

\begin{table}[htb!]
\begin{center}

\caption{The systematic uncertainties from the Monte Carlo simulation, the
fitting condition, and total for each mode are shown.}
\vspace{0.5cm}
\label{system}

Systematic Error 

\begin{tabular}{|c | c | c | c| c | }
\hline
  Mode            & Simulation & Fit     &  Total   \\ 
\hline

$\frac{\Gamma (\Xi _{c}^{+}\! ~\rightarrow~ \! \Sigma ^{+}K^{-}\pi ^{+})}{\Gamma
(\Xi_{c}^{+} \! ~\rightarrow~ \! \Xi^{-}\pi^+ \pi^+)}$ &
 0.00& 0.04& 0.04\\ 
\hline

$\frac{\Gamma (\Xi_{c}^{+} \! ~\rightarrow~ \! \Sigma^{+} \bar{K}^{*}(892)^0)}{\Gamma (\Xi_{c}^{+} \! ~\rightarrow~ \!
 \Xi^{-} \pi^{+} \pi ^{+}) }$  &
 0.00& 0.06& 0.06\\ 

\hline

$\frac{\Gamma (\Xi_{c}^{+} \! ~\rightarrow~ \! \Sigma ^{+}K^{+}K^{-})}{\Gamma(\Xi _{c}^{+} \! ~\rightarrow~ \!
\Sigma^{+} K ^{-}\pi ^{+}) }$  &
 ---&0.01&0.01\\ 

\hline

$\frac{\Gamma (\Xi_{c}^{+}\! ~\rightarrow~ \! \Lambda ^{0} K^- \pi^+ \pi^+)}{\Gamma (\Xi_{c}^{+}\! ~\rightarrow~ \!
\Xi ^{-}\pi ^{+}\pi ^{+}) }$  &
 0.05& 0.04 & 0.06\\ 

\hline

$\frac{\Gamma (\Xi _{c}^{+}\! ~\rightarrow~ \! \Omega ^{-} K^+ \pi^+)}{\Gamma (\Xi _{c}^{+}\! ~\rightarrow~ \!
\Xi ^{-}\pi ^{+}\pi ^{+}) }$  &
 0.03&0.01&0.03\\ 

\hline

$\frac{\Gamma (\Xi _{c}^{+}\! ~\rightarrow~ \! \Sigma^{*}(1385)^+ \bar{K}^0)}{\Gamma (\Xi _{c}^{+}\! ~\rightarrow~ \!
\Xi ^{-}\pi ^{+}\pi ^{+}) }$  &
0.19 & 0.14 & 0.24\\ 

\hline



\hline
\hline
\end{tabular}
\end{center}
\end{table}

\begin{table}[htb!]
\begin{center}

\caption{FOCUS results compared to previous measurements. The
relative efficiencies are computed with respect to the normalization mode
(for $\Xi_{c}^{+}\rightarrow \Xi^*(1690)^0 K^+$ we do not correct for
the branching fraction of $\Xi^*(1690)^0\rightarrow \Sigma^+ K^-$ as it is not known).}
\vspace{0.5cm}
\label{tb:results}
Relative Branching Ratio

\begin{tabular}{|c | c | c | c| c | }
\hline\hline


Decay Mode               & \shortstack{Efficiency\\Ratio}      & FOCUS     &  CLEO & SELEX   \\ \hline

$\frac{\Gamma (\Xi _{c}^{+}\! ~\rightarrow~ \! \Sigma ^{+}K^{-}\pi ^{+})}{\Gamma
(\Xi_{c}^{+} \! ~\rightarrow~ \! \Xi^{-}\pi^+ \pi^+)}$	  & 1.04 & 
$0.91 \pm 0.11 \pm 0.04 $ & $1.18 \pm 0.26 \pm 0.17 $  & $0.92 \pm 0.20 \pm 0.07$ \\ \hline

$\frac{\Gamma (\Xi_{c}^{+} \! ~\rightarrow~ \! \Sigma^{+} \bar{K}^{*}(892)^0)}{\Gamma (\Xi_{c}^{+} \! ~\rightarrow~ \!
 \Xi^{-} \pi^{+} \pi ^{+}) }$ & 0.57 & $0.78 \pm 0.16 \pm 0.06 $  & $0.92 \pm 0.27 \pm 0.17$ & ---\\ \hline

$\frac{\Gamma (\Xi_{c}^{+} \! ~\rightarrow~ \! \Sigma ^{+}K^{+}K^{-})}{\Gamma(\Xi _{c}^{+} \! ~\rightarrow~ \!
\Sigma^{+} K ^{-}\pi ^{+}) }$ & 0.77 & $0.16 \pm 0.06 \pm 0.01$  & --- & --- \\ \hline

$\frac{\Gamma (\Xi_{c}^{+} \! ~\rightarrow~ \! \Sigma ^{+}\phi)}{\Gamma(\Xi _{c}^{+} \! ~\rightarrow~ \!
\Sigma^{+} K ^{-}\pi ^{+}) }$ & 0.33 &$<0.12$ at 90$\%$ C.L.& --- & --- \\ \hline

$\frac{\Gamma (\Xi_{c}^{+} \! ~\rightarrow~ \! \Xi^*(1690)^0 K^+)}{\Gamma(\Xi _{c}^{+} \! ~\rightarrow~ \!
\Sigma^{+} K ^{-}\pi ^{+}) }$ & 0.57 &$<0.05$ at 90$\%$ C.L.& --- & --- \\ \hline

$\frac{\Gamma (\Xi_{c}^{+}\! ~\rightarrow~ \! \Lambda ^{0} K^- \pi^+ \pi^+)}{\Gamma (\Xi_{c}^{+}\! ~\rightarrow~ \!
\Xi ^{-}\pi ^{+}\pi ^{+}) }$ & 1.09 & $0.28\pm 0.06\pm 0.06$  & $0.58\pm 0.16\pm 0.07$ & ---  \\\hline

$\frac{\Gamma (\Xi _{c}^{+}\! ~\rightarrow~ \! \Omega ^{-} K^+ \pi^+)}{\Gamma (\Xi _{c}^{+}\! ~\rightarrow~ \!
\Xi ^{-}\pi ^{+}\pi ^{+}) }$ & 1.40 & $0.07 \pm 0.03 \pm 0.03 $  & --- & ---\\ 
 &  & $<0.12$ at 90$\%$ C.L.   & & \\ \hline

$\frac{\Gamma (\Xi _{c}^{+}\! ~\rightarrow~ \! \Sigma^{*}(1385)^+ \bar{K}^0)}{\Gamma (\Xi _{c}^{+}\! ~\rightarrow~ \!
\Xi ^{-}\pi ^{+}\pi ^{+}) }$ & 0.21 & $1.00 \pm 0.49 \pm 0.24 $  & --- & ---\\ 

 &  & $<1.72$ at 90$\%$ C.L.   & & \\ \hline

$\frac{\Gamma (\Xi _{c}^{+}\! ~\rightarrow~ \! \Xi^{*}(1530)^0 \pi ^{+})}{\Gamma
(\Xi_{c}^{+} \! ~\rightarrow~ \! \Xi^{-}\pi^+ \pi^+)}$	  & 0.62 & $<0.10 $ at 90$\%$ C.L. & $<0.2 $ at 90$\%$ C.L & ---\\ \hline

\hline
\hline
\end{tabular}
\end{center}
\end{table}

\section{Conclusions}
We investigate and measure the relative branching
ratios of several decay modes of the charm baryon $\Xi _{c}^{+}$.
We report the first evidence for the Cabibbo suppressed decay $\Xi _{c}^{+}\rightarrow \Sigma ^{+}K^{+}K^{-}$
and we investigate the contribution from the resonant modes $\Xi _{c}^{+}\rightarrow \Sigma ^{+}\phi$
and $\Xi _{c}^{+}\rightarrow \Xi^*(1690)^0 K^{+}$.
We report an indication of the decays $\Xi _{c}^{+}\rightarrow \Omega ^{-}K^{+}\pi ^{+}$ and
$\Xi _{c}^{+}\rightarrow \Sigma^*(1385)\bar{K}^0$.
We also report improved measurements of $\Xi _{c}^{+}$ decays in
the final state $\Sigma ^{+}K^{-}\pi ^{+}$, $\Sigma ^{+}\bar{K}^{*}(892)^{0}$
and $\Lambda ^{0}K^{-}\pi ^{+}\pi ^{+}$. These last three results
agree with previous measurements from the
the CLEO and SELEX collaborations. Finally, we report an improved measurement 
of the limit for the resonant decay 
$\Xi _{c}^{+}\rightarrow \Xi^{*}(1530)^0 \pi ^{+}$.

\section{Acknowledgements}
We wish to acknowledge the assistance of the staffs of Fermi National
Accelerator Laboratory, the INFN of Italy, and the physics departments of the
collaborating institutions. This research was supported in part by the U.~S.
National Science Foundation, the U.~S. Department of Energy, the Italian
Istituto Nazionale di Fisica Nucleare and Ministero dell'Universit\`a e della
Ricerca Scientifica e Tecnologica, the Brazilian Conselho Nacional de
Desenvolvimento Cient\'{\i}fico e Tecnol\'ogico, CONACyT-M\'exico, the Korean
Ministry of Education, and the Korean Science and Engineering Foundation.

\bibliographystyle{myapsrev}
\bibliography{main}

\end{document}